\documentclass{elsart}
\usepackage{epsfig,amssymb}
\journal{Astroparticle Physics}
\begin{document}
\begin{frontmatter}

\title{First results of the Instrumentation Line for the deep-sea
ANTARES neutrino telescope}

\begin{abstract}
In 2005, the ANTARES Collaboration deployed and operated at a depth of
2500~m a so-called Mini Instrumentation Line equipped with Optical
Modules (MILOM) at the ANTARES site. The various data acquired during
the continuous operation from April to December 2005 of the MILOM
confirm the satisfactory performance of the Optical Modules, their
front-end electronics and readout system, as well as the calibration
devices of the detector. The in-situ measurement of the Optical Module
time response yields a resolution better than 0.5~ns. The performance
of the acoustic positioning system, which enables the spatial
reconstruction of the ANTARES detector with a precision of about
10~cm, is verified. These results demonstrate that with the full
ANTARES neutrino telescope the design angular resolution of better
than $0.3^\circ$ can be realistically achieved.
\end{abstract}

\renewcommand{\thefootnote}{\fnsymbol{footnote}}
\begin{center}
\author[IFIC]{J.A.~Aguilar},
\author[Mulhouse]{A.~Albert},
\author[Rome]{F.~Ameli},
\author[Genova]{M.~Anghinolfi},
\author[Erlangen]{G.~Anton},
\author[Saclay]{S.~Anvar},
\author[CPPM]{E.~Aslanides},
\author[CPPM]{J-J.~Aubert},
\author[Bari]{E.~Barbarito},
\author[LAM]{S.~Basa}
% ...
\footnotetext[1]{Deceased.}
\newpage
\nopagebreak[3]
\address[APC]{APC -- AstroParticule et Cosmologie, UMR 7164 (CNRS, Universit\'e Paris 7, CEA, Observatoire
de Paris), 11, place Marcelin Berthelot, 75231 Paris Cedex 05, France}
\vspace*{-0.40\baselineskip}
\nopagebreak[3]
\address[Bari]{Dipartimento Interateneo di Fisica e Sezione INFN, Via E. Orabona 4, 70126 Bari, Italy}
\vspace*{-0.40\baselineskip}
\nopagebreak[3]
\address[Bologna]{Dipartimento di Fisica dell'Universit\`a e Sezione INFN, Viale Berti Pichat 6/2, 40127 Bologna, Italy}
\vspace*{-0.40\baselineskip}
\nopagebreak[3]
\address[COM]{COM -- Centre d'Oc\'eanologie de Marseille, CNRS/INSU et Universit\'e de la
M\'editerran\'ee, 163 Avenue de Luminy, Case 901, 13288 Marseille Cedex 9, France}
\vspace*{-0.40\baselineskip}
\nopagebreak[3]
\address[CPPM]{CPPM -- Centre de Physique des Particules de Marseille, CNRS/IN2P3 et Universit\'e de la
M\'editerran\'ee, 163 Avenue de Luminy, Case 902, 13288 Marseille Cedex 9, France}
\vspace*{-0.40\baselineskip}
\nopagebreak[3]
\address[Catania]{Dipartimento di Fisica ed Astronomia dell'Universit\`a e Sezione INFN, Viale Andrea Doria 6, 95125 Catania, Italy}
\vspace*{-0.40\baselineskip}
\nopagebreak[3]
\address[Erlangen]{Friedrich-Alexander-Universit\"at Erlangen-N\"urnberg, Physikalisches Institut, Erwin-Rommel-Str.\ 1, D-91058 Erlangen, Germany}
\vspace*{-0.40\baselineskip}
\nopagebreak[3]
\address[GEOAZUR]{G\'eoSciences Azur, CNRS/INSU, IRD, Universit\'e de Nice Sophia-Antipolis, Universit\'e
Pierre et Marie Curie -- Observatoire Oc\'eanologique de Villefranche, BP48, 2 quai de la Darse, 06235 Villefranche-sur-Mer Cedex, France}
\vspace*{-0.40\baselineskip}
\nopagebreak[3]
\address[Genova]{Dipartimento di Fisica dell'Universit\`a e Sezione INFN, Via Dodecaneso 33, 16146 Genova, Italy}
\vspace*{-0.40\baselineskip}
\nopagebreak[3]
\address[IFIC]{IFIC -- Instituto de F\'{\i}sica Corpuscular, Edificios Investigaci\'on de Paterna, CSIC -- Universitat de Val\`encia, Apdo. de Correos 22085, 46071 Valencia, Spain}
\vspace*{-0.40\baselineskip}
\nopagebreak[3]
\address[IFREMER/Brest]{IFREMER -- Centre de Brest, BP 70, 29280 Plouzan\'e, France}
\vspace*{-0.40\baselineskip}
\nopagebreak[3]
\address[IFREMER/Toulon]{IFREMER -- Centre de Toulon/La Seyne Sur Mer, Port Br\'egaillon, Chemin Jean-Marie Fritz, 83500, La Seyne sur Mer, France}
\vspace*{-0.40\baselineskip}
\nopagebreak[3]
\address[INAF]{INAF-IASF, via P. Gobetti 101, 40129 Bologna, Italy}
\vspace*{-0.40\baselineskip}
\nopagebreak[3]
\address[IPHC]{IPHC -- Institut Pluridisciplinaire Hubert Curien, Universit\'e Louis Pasteur (Strasbourg
1) et IN2P3/CNRS, 23 rue du Loess, BP 28, 67037 Strasbourg Cedex 2, France}
\vspace*{-0.40\baselineskip}
\nopagebreak[3]
\address[ISS]{Institute for Space Sciences, 77125 Bucharest, Magurele, Romania}
\vspace*{-0.40\baselineskip}
\nopagebreak[3]
\address[ITEP]{ITEP -- Institute for Theoretical and Experimental Physics, B.~Cheremushkinskaya 25, 117259 Moscow, Russia}
\vspace*{-0.40\baselineskip}
\nopagebreak[3]
\address[KVI]{Kernfysisch Versneller Instituut (KVI), University of Groningen, Zernikelaan 25, 9747 AA Groningen, The Netherlands}
\vspace*{-0.40\baselineskip}
\nopagebreak[3]
\address[LAM]{LAM -- Laboratoire d'Astrophysique de Marseille, CNRS/INSU et Universit\'e de Provence, Traverse du Siphon -- Les Trois Lucs, BP 8, 13012 Marseille Cedex 12, France}
\vspace*{-0.40\baselineskip}
\nopagebreak[3]
\address[LNS]{INFN -- Labaratori Nazionali del Sud (LNS), Via S. Sofia 44, 95123 Catania, Italy}
\vspace*{-0.40\baselineskip}
\nopagebreak[3]
\address[Leeds]{School of Physics \& Astronomy, University of Leeds LS2 9JT, UK}
\vspace*{-0.40\baselineskip}
\nopagebreak[3]
\address[Mulhouse]{GRPHE -- Groupe de Recherche en Physique des Hautes Energies, Universit\'e de Haute Alsace, 61 Rue Albert Camus, 68093 Mulhouse Cedex, France}
\vspace*{-0.40\baselineskip}
\nopagebreak[3]
\address[NIKHEF]{Nationaal Instituut voor Kernfysica en Hoge-Energiefysica (NIKHEF), Kruislaan 409, 1098 SJ Amsterdam, The Netherlands}
\vspace*{-0.40\baselineskip}
\nopagebreak[3]
\address[Pisa]{Dipartimento di Fisica dell'Universit\`a e Sezione INFN, Largo B.~Pontecorvo 3, 56127 Pisa, Italy}
\vspace*{-0.40\baselineskip}
\nopagebreak[3]
\address[Rome]{Dipartimento di Fisica dell'Universit\`a "La Sapienza" e Sezione INFN, P.le Aldo Moro 2, 00185 Roma, Italy}
\vspace*{-0.40\baselineskip}
\nopagebreak[3]
\address[Saclay]{DSM/Dapnia -- Direction des Sciences de la  Mati\`ere, laboratoire de recherche sur les
lois fondamentales de l'Univers, CEA Saclay, 91191 Gif-sur-Yvette Cedex, France}
\vspace*{-0.40\baselineskip}
\nopagebreak[3]
\address[Sheffield]{Dept.\ of Physics and Astronomy, University of Sheffield, Sheffield S3 7RH, UK}
\vspace*{-0.40\baselineskip}
\nopagebreak[3]
\address[UU]{Universiteit Utrecht, Faculteit Betawetenschappen, Princetonplein 5, 3584 CC Utrecht, The Netherlands}
\vspace*{-0.40\baselineskip}
\nopagebreak[3]
\address[UvA]{Universiteit van Amsterdam, Instituut voor Hoge-Energiefysica, Kruislaan 409, 1098 SJ Amsterdam, The Netherlands}
\vspace*{-0.40\baselineskip}
\end{center}
\end{frontmatter}

\section{Introduction}

The ANTARES Collaboration is building a large underwater neutrino
telescope located at a depth of 2500~m in the Mediterranean Sea,
offshore from Toulon in France~\cite{proposal}. The experiment aims to
detect neutrinos with energies above 10~GeV by means of the Cherenkov
light emitted in sea water by charged particles produced in
neutrino interactions with the surrounding medium. Photons are recorded
by a lattice of Optical Modules~\cite{OM}, consisting of 10''
hemispherical photomultiplier tubes~\cite{PM} housed in pressure
resistant glass spheres, installed along a set of mooring lines. The
ANTARES detector will consist of 12 lines of 25 storeys, each storey
being equipped with a triplet of Optical Modules and an electronic
container mounted on a titanium frame, giving thus a grand total of
900 OMs. Some storeys also support a hydrophone for acoustic
positioning or an LED Optical Beacon used for time
calibration. Every line is individually connected to a Junction Box by
an interconnecting cable of a few hundred metres long, laid down on the
sea bed. The Junction Box is itself linked to the shore station by a
40~km long electro-optical cable equipped with 48 optical
fibres. Details of the ANTARES detector architecture, components and
calibration devices will be found in forthcoming publications. The
installation of the 12 lines of the detector has started in February
2006 with completion expected in 2007.

In Spring 2003, two prototype lines, the Prototype Sector Line (PSL)
equipped with 15 Optical Modules (OMs) and the Mini Instrumentation
Line (MIL) hosting calibration and environmental measurement devices,
were deployed, connected and operated for a few months at the ANTARES
site. This operation allowed a demonstration of the main aspects of
the design of the detector and measurements of the background counting
rates in the OMs, due to bioluminescence and $^{40}$K decays, over a
period of about four months~\cite{TAUP}. After the experience of the
PSL and MIL lines, significant changes were made to the detector
design. In parallel to the launching of the mass production of all
detector elements, the ANTARES Collaboration built a new version of
the Mini Instrumentation Line based on the final design of all
electronics and mechanics. This line, which also includes an
extra storey with three OMs, has been named the MILOM. The main
objective of the MILOM operation is to provide an in-situ check of the
modified detector elements and a validation of the performance of the
time calibration and the acoustic positioning devices. It was also an
excellent opportunity to validate the tools and procedures used during
the integration and deployment of a line before their application to
the first complete ANTARES detector line. Furthermore, the MILOM
houses environmental instruments needed for the calibration of the
detector and the monitoring of the water physical properties.

Completed in December 2004, the MILOM line was deployed on the ANTARES
site~\cite{LightTrans,fouling}, located at $42^\circ48'$\,N\,--\,$6^\circ10'$\,E, in March 2005 and connected
to the existing Junction Box with the Remote Operated Vehicle (ROV) Victor
of IFREMER in April 2005. It has been operating since then with almost
permanent shift crews from the ANTARES shore station located in La
Seyne-sur-Mer. This paper highlights the main results obtained from
the MILOM operation in 2005, in particular the analysis of the Optical
Module signals and of the calibration device data.

\begin{figure}[htb]
\begin{center}
\epsfig{file=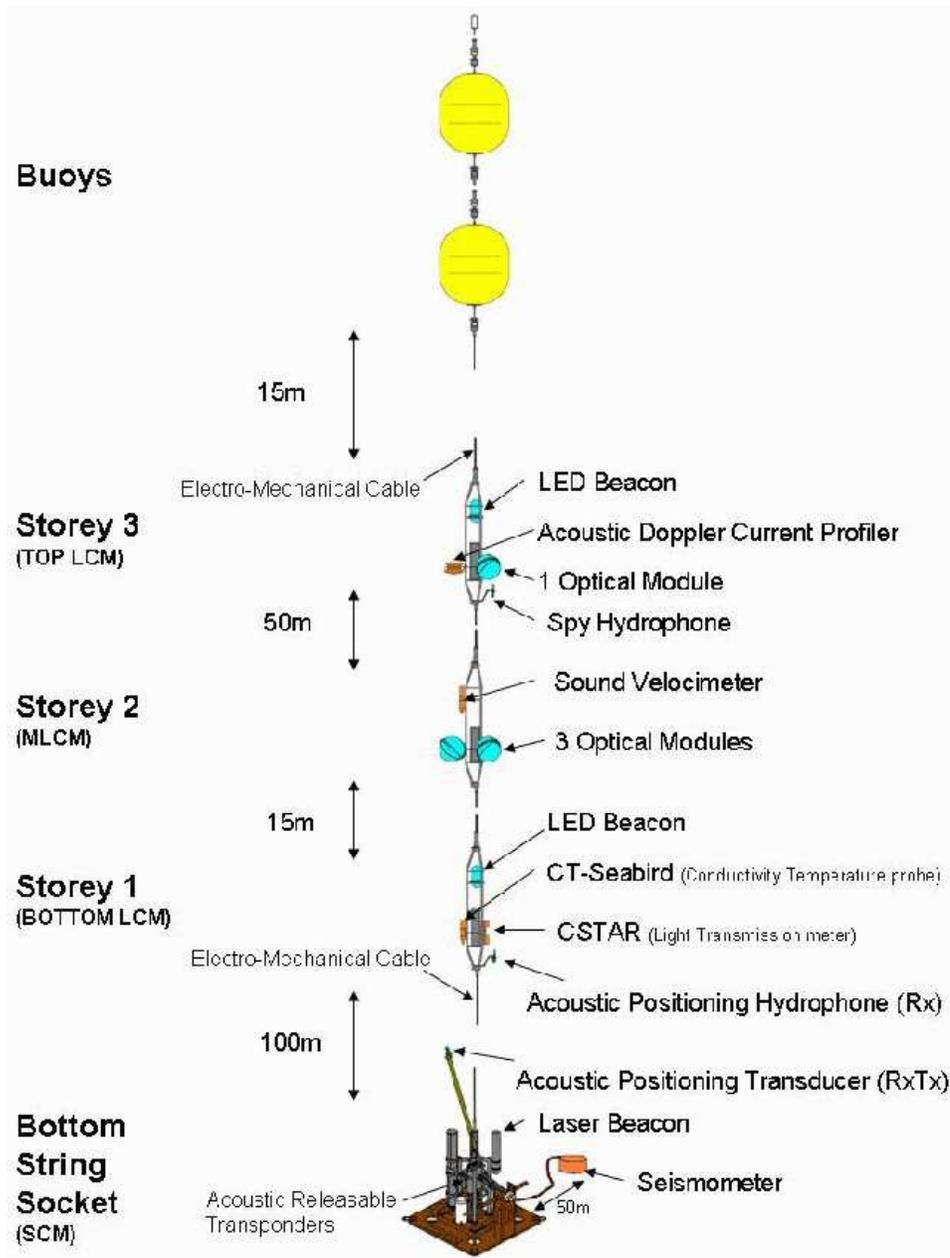,width=0.9\textwidth}
\caption{Layout of the MILOM line.}
\label{MILOM layout}
\end{center}
\end{figure}

\section{The MILOM line}

As shown in figure~\ref{MILOM layout}, the MILOM consists of an
instrumented releasable anchor, the Bottom String Socket (BSS), and of
three storeys located respectively at 100~m, 117~m and 169~m above the
sea bed (the inter-storey spacings indicated in figure~\ref{MILOM
layout} correspond to the cable lengths). The line is maintained in an
almost vertical position by two buoys located at the top.

The MILOM is equipped with four Optical Modules: a triplet of OMs on
the second storey, as for a standard ANTARES optical line storey, and
a single additional OM fixed on the top storey. The line also supports
three intense light sources used mainly for the OM timing calibration:
the Laser Beacon located on the BSS and two LED Optical Beacons
attached to the bottom and top storey, respectively. In order to allow
the reconstruction of the line shape geometry, the MILOM is equipped
with biaxial tiltmeters and compasses located in the electronics
container of every storey, and with two acoustic positioning modules:
an emission/reception (RxTx) module with its transducer on the BSS and
a reception (Rx) module with its hydrophone on the bottom storey. In
addition, the MILOM hosts various environmental devices: an acoustic
current profiler\footnote{300 kHz direct reading ADCP Workhorse Monitor from 
Teledyne RD Instruments, http://www.rdinstruments.com/monitor.html}
monitors the intensity and direction of the underwater flow; 
a sound velocimeter\footnote{Sound velocimeter Ref QUUX-3A(A)
from Genisea/ECA, \\http://perso.wanadoo.fr/genisea/sound\_velocimeter.htm}
records the local value of the sound velocity; 
a CT probe\footnote{MicroCAT C-T sensor SBE~37-SI from Sea Bird Electronics,\\
http://www.seabird.com/products/spec\_sheets/37sidata.htm} measures the
conductivity and temperature of the sea water; a transmission
meter\footnote{C-Star transmissiometer from WET Labs,\\
http://www.wetlabs.com/products/cstar/cstar.htm}
monitors the light attenuation of the water; a Spy
Hydrophone records the acoustic activity from the positioning beacons,
surface or biological noise; and a broadband 
seismometer\footnote{Triaxial broadband seismometer CMG-3T from G\"uralp Systems,\\
http://www.guralp.net/products/}
installed in the sea bed sediment 50~m away from the MILOM anchor
which monitors the seismic activity at the site. Finally, the MILOM
BSS is equipped with two releasable autonomous
transponders\footnote{Acoustic releasable transponder RT661B2T from IXSEA,\\ 
http://www.ixsea.com/php/contenu/en/p-oceano\_ocean.php} 
which enable the monitoring of the line
anchor position during its deployment and allow the release of the BSS
from its dead weight in order to recover the line.

All instruments deliver their data in real time and can be remotely
controlled from the ANTARES shore station through a Gb Ethernet
network. Every storey is equipped with a Local Control Module (LCM)
which contains the electronics boards for the OM signal processing, the
instrument readout, the acoustic positioning, the power system and the
data transmission. On the middle storey, the Master Local Control
Module (MLCM) also contains an Ethernet switch board, which
multiplexes the data acquisition (DAQ) channels from the other
storeys. At the bottom of the line, the BSS is equipped with a String
Control Module (SCM) which contains the local readout and DAQ
electronics, as well as the power system for
the whole line. Finally, both MCLM and SCM include a Dense Wavelength
Division Multiplexing system used for data transmission in order to
merge several 1Gb/s Ethernet channels on the same pair of optical
fibres by using different laser wavelengths. Although a local trigger
requiring time coincidences between OMs of the same storey can be
activated in each LCM, most of the time the large bandwidth 
of the DAQ system allows the transmission of all recorded 
OM signals to shore. A dedicated computer farm can then perform a
global selection of the OM hits of the interesting physics events
from the data recorded by the whole detector.

While the recording of the OM signals and instrument data of the MILOM
started immediately after its connection, the first months of
operation have been dedicated to online software development and
tuning of the detector settings. The smooth data taking of the MILOM
started in September 2005.

\begin{figure}[htb]
\begin{center}
\epsfig{file=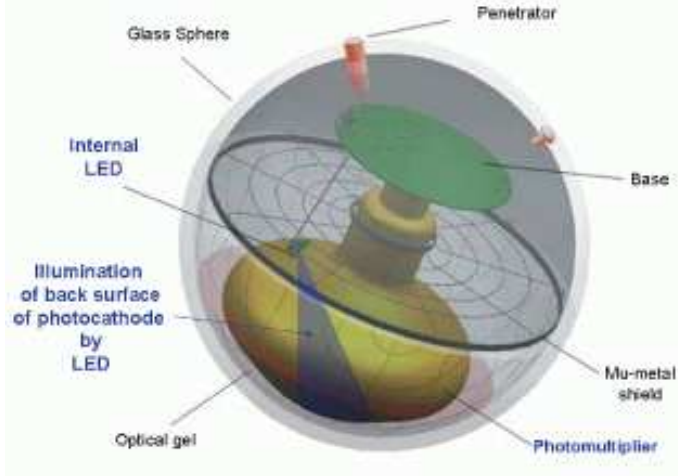,width=0.65\textwidth}
\caption{Sketch of the ANTARES Optical Module (for a complete
description, see ref~\cite{OM}).}
\label{figOM}
\end{center}
\end{figure}

\section{The Optical Module data} 

The ANTARES Optical Module consists of a 10'' Hamamatsu
photomultiplier tube (PMT) housed in a pressure resistant glass sphere (see
figure~\ref{figOM}). The 900 PMTs foreseen for the ANTARES detector
have been selected and fully characterized to work with a threshold
below the single photo-electron level with a mean transit time spread (TTS)
of $\sigma\sim1.3$~ns (FWHM $\sim3.0$~ns)~\cite{PM}. The PMT signal is processed
by the Analogue Ring Sampler (ARS) ASIC which measures the arrival
time and charge of the pulse~\cite{ARS}. Only this information is
sent to shore in the case where it is compatible with a single
photo-electron (SPE) pulse, while the ARS can perform a full
digitisation of the PMT signal for larger amplitudes. In order to
minimize the dead time, every OM is read out by a pair of ARS chips
which treat the signal alternately according to a token ring protocol.

\begin{figure}[htb]
\begin{center}
\epsfig{file=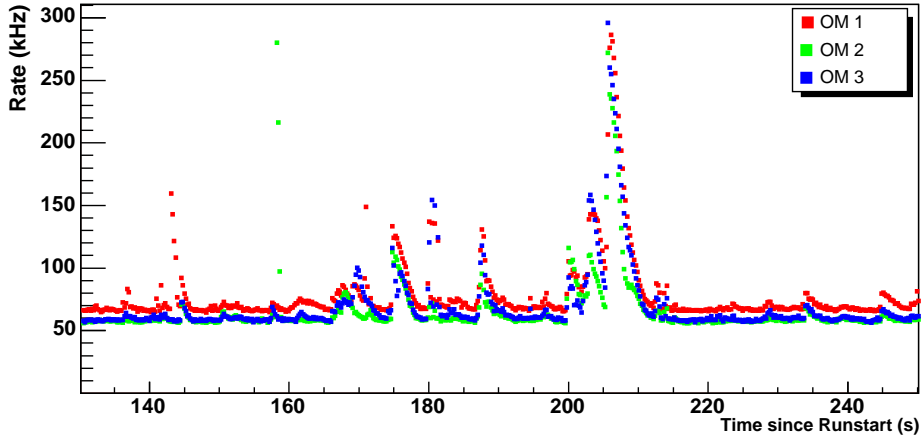,width=1.0\textwidth}
\caption{Example of counting rates for the three OMs located on the
second storey of the MILOM over a period of 120 seconds.}
\label{Rates}
\end{center}
\end{figure}

During normal operation, the PMT high voltage is set so as to obtain a
gain of $5\times10^7$ leading to a SPE signal amplitude of about
45~mV. The readout trigger threshold of the ARS is set to
$\sim0.5$~photo-electrons. Figure~\ref{Rates} shows an example of the
counting rates recorded by the three OMs located on the MILOM second
storey over a period of 120 seconds. The counting rates exhibit a
baseline largely dominated by optical background due to $^{40}$K
decays and bioluminescence coming from bacteria, as well as bursts of
a few seconds duration produced by bioluminescent emission of 
macro-organisms~\cite{biolum}. The fourth OM located on the top storey has
not worked since the beginning of the MILOM operation because it shares
its power supply with a faulty Optical Beacon placed on that same
storey. A visual survey of the MILOM line with the ROV Victor
revealed that this Optical Beacon is full of water due to a leak.

\begin{figure}[htb]
\begin{center}
\epsfig{file=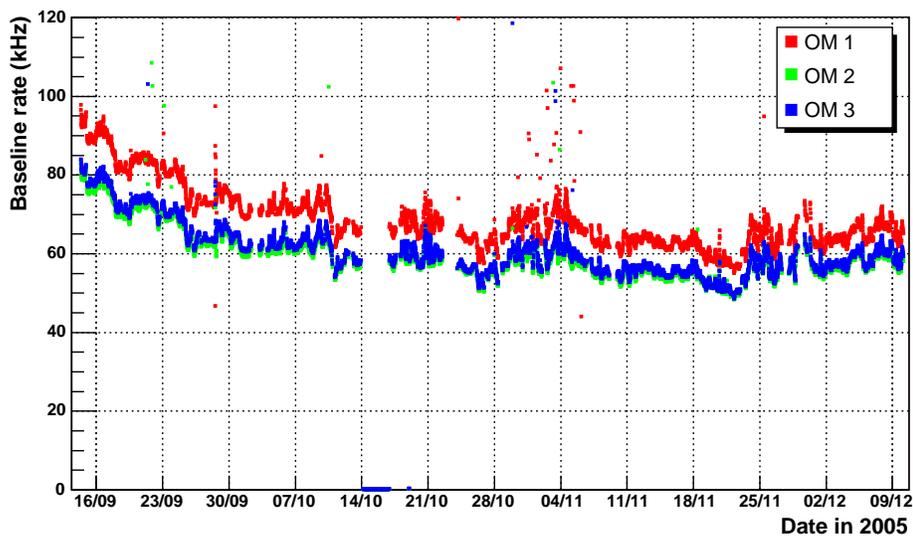,width=1.0\textwidth}
\caption{Baseline rates for the three OMs during Autumn 2005.}
\label{Baselines}
\end{center}
\end{figure}

For each OM, the baseline rate is defined as the average of the
minimum counting rate during periods of 15
minutes. Figure~\ref{Baselines} shows the summary of the baseline
rates extracted from the data recorded by the three OMs for a period
of three months during Autumn 2005. The variability of the
bioluminescence component of the baseline, already observed during the
operation of the Prototype Sector Line in 2003, is confirmed by the
present data. The baseline counting rates decrease
from 80-90~kHz in mid-September to about 60~kHz in November
2005. Figures~\ref{Rates} and~\ref{Baselines} also clearly reveal that
the counting rate of one Optical Module (OM1) is systematically larger
by about 15\% with respect to the other two. This
difference 
is attributed to a lower threshold value of the PMT.

\begin{figure}[htb]
\begin{center}
\epsfig{file=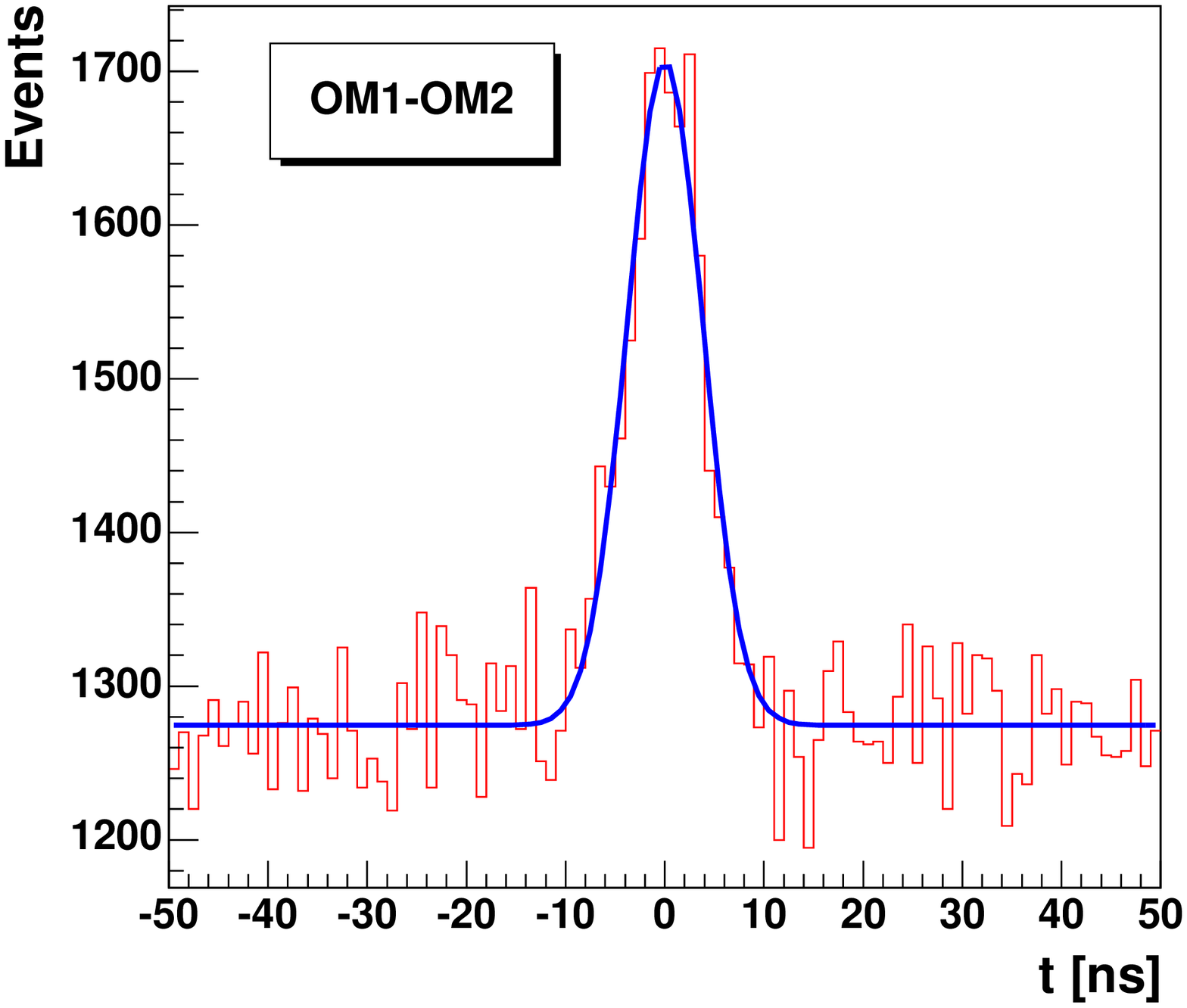,width=0.32\textwidth}
\epsfig{file=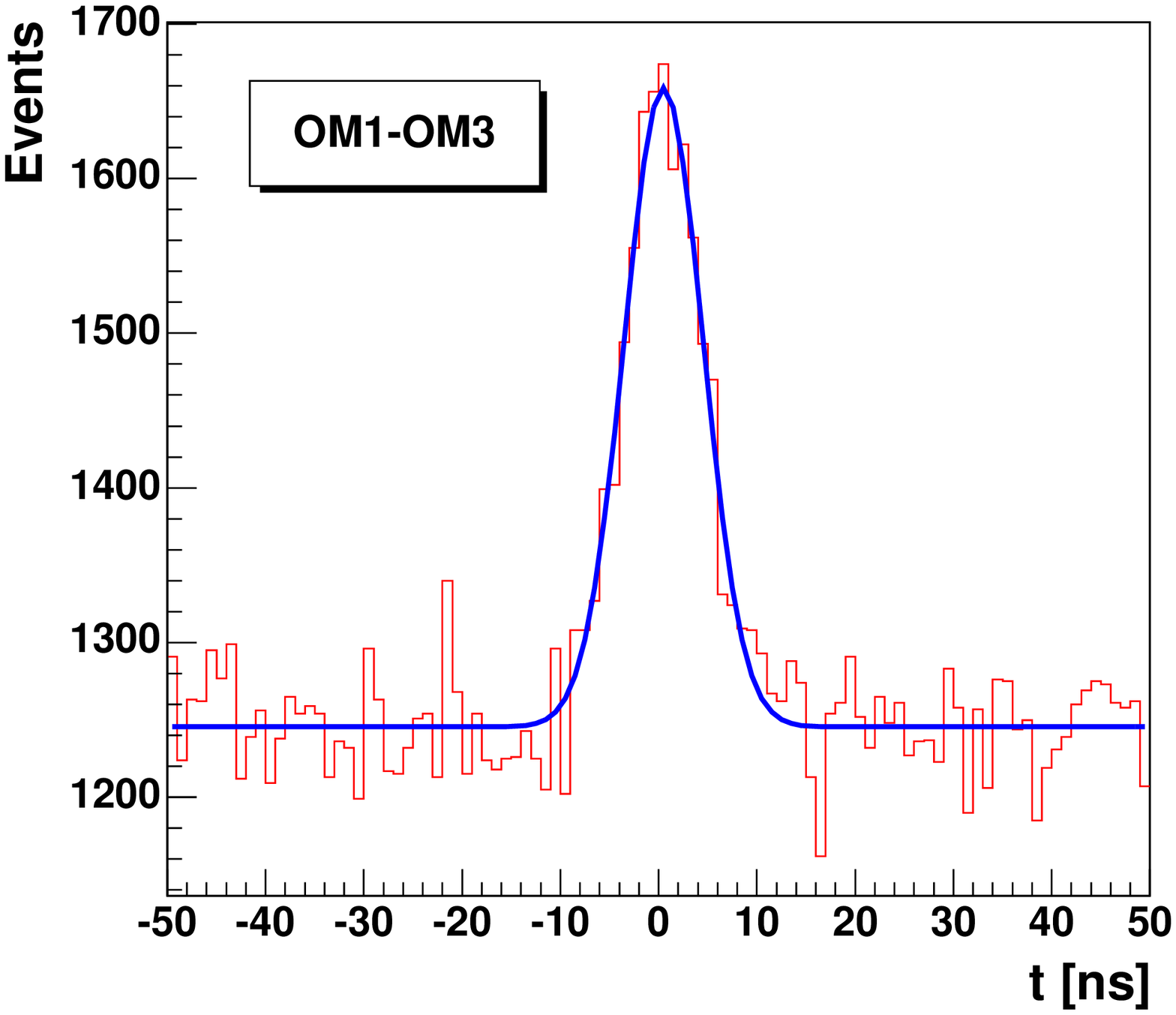,width=0.32\textwidth}
\epsfig{file=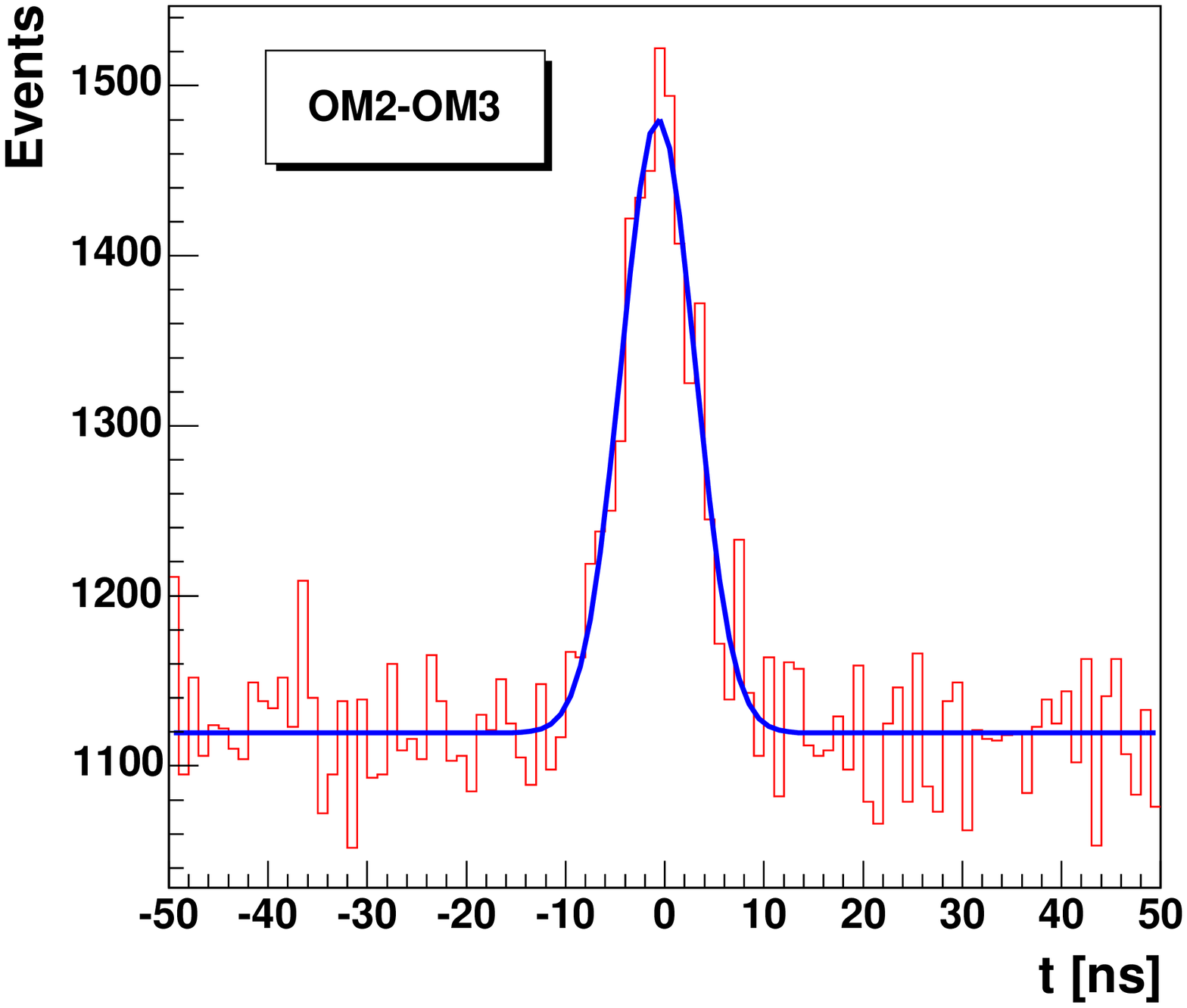,width=0.32\textwidth}
\caption{Distributions of time differences between hits registered by
the three OM pairs. The solid lines are fits of a Gaussian and a constant 
background to these distributions used to extract the $^{40}$K coincidence rates.}
\label{Coinc}
\end{center}
\end{figure}

Time coincidences between signals of OM pairs have also been
studied. Figure~\ref{Coinc} shows typical distributions of the time
delay between signals from the three possible combinations. The
distributions show peaks due to $^{40}$K radioactive decays producing 
two detected photons, superimposed upon a flat background of
random coincidences. The genuine $^{40}$K coincidence rates obtained from
fits to these distributions are 
$13.0\pm0.5$~Hz for
the OM1-OM2 and OM1-OM3 pairs and $10.5\pm0.4$~Hz for the OM2-OM3 pair.
These results are found to be stable within the statistical errors over a
period of four months. The measured rates are in good agreement with a simulation of the
signals induced by $^{40}$K decays which leads to a coincidence
rate of 12~Hz with a 4~Hz systematic error due to uncertainties in the
effective area and angular response of the OMs.  These measurements
also confirm that the larger counting rate observed with OM1 is not
due to noise in its front-end electronics.

\begin{figure}[htb]
\begin{center}
\epsfig{file=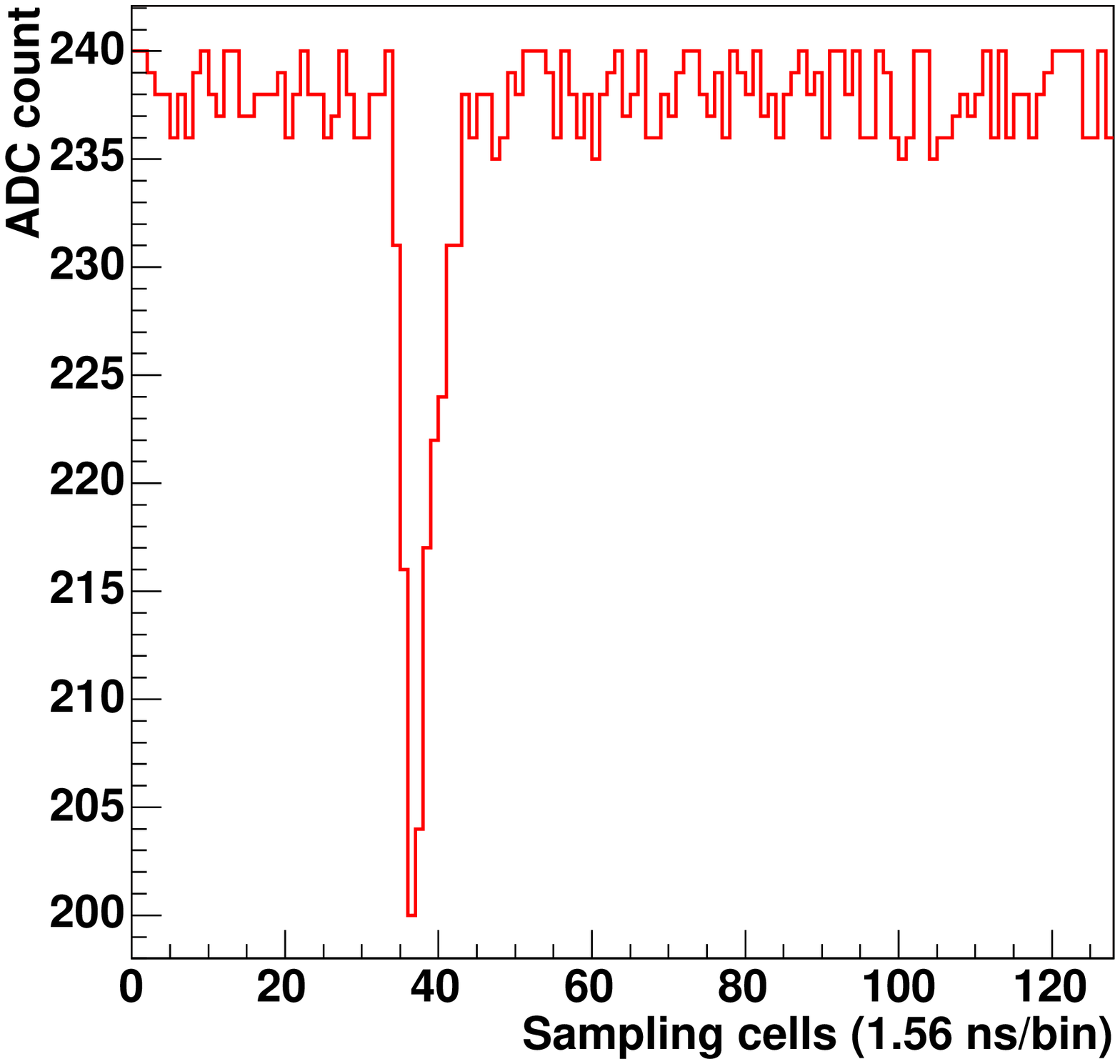,width=0.32\textwidth}
\epsfig{file=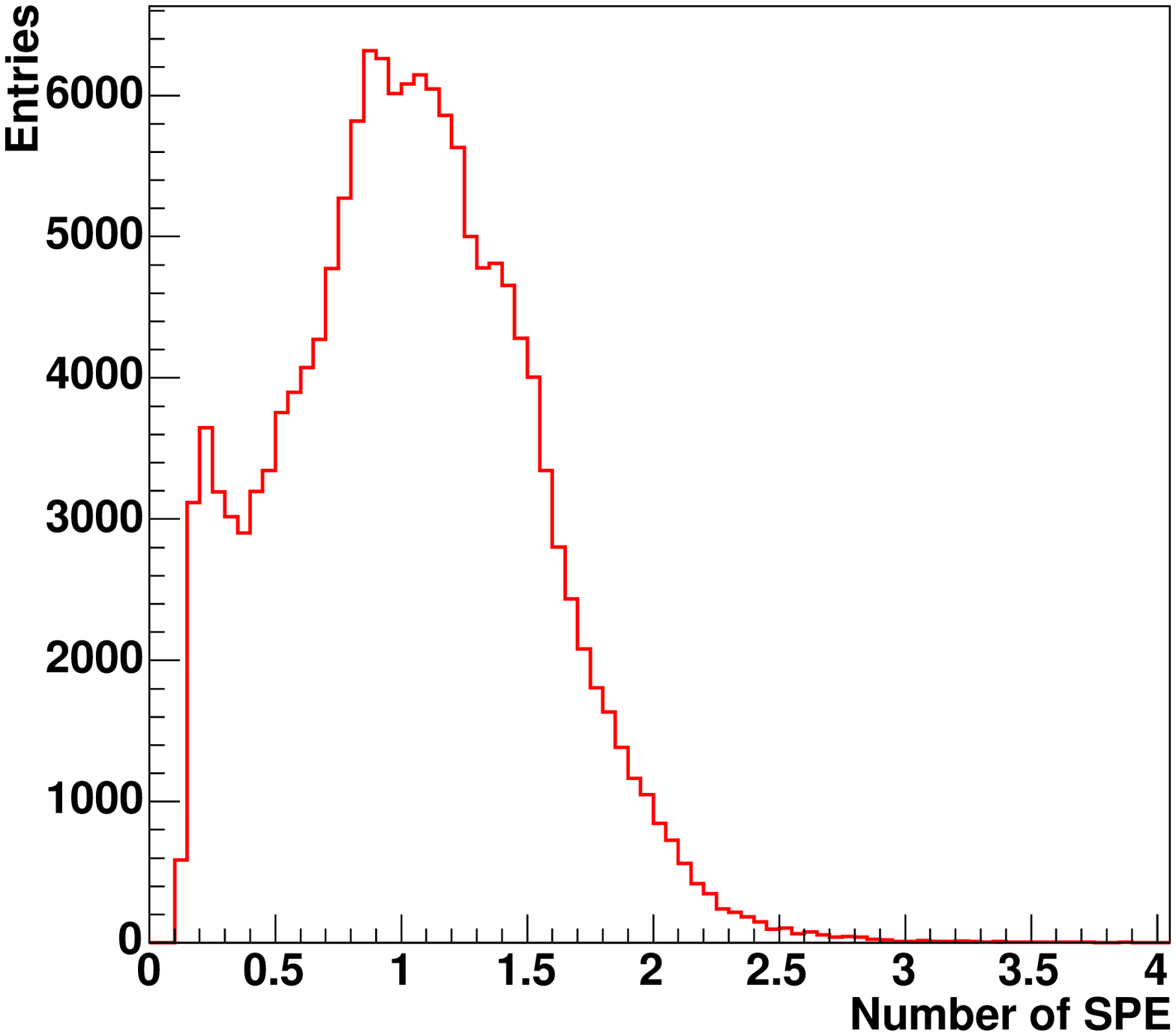,width=0.32\textwidth}
\epsfig{file=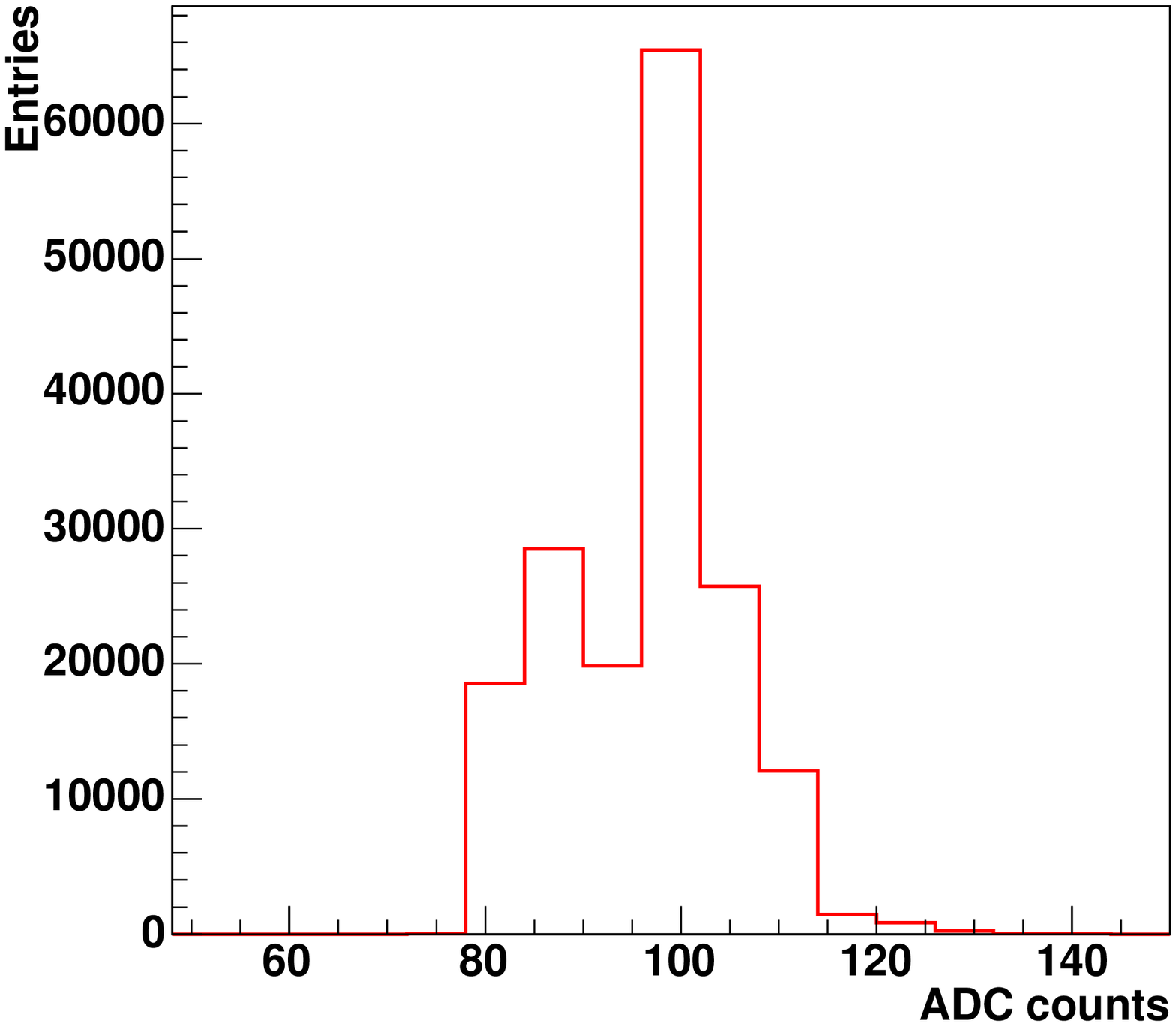,width=0.32\textwidth}
\caption{Example of a waveform sampling of an OM signal
(left). Charge distribution of the PMT signal obtained by integrating
the WF samples (middle) and measured by the AVC circuit of the ARS
(right).}
\label{AVC}
\end{center}
\end{figure}

As mentioned previously, the ARS has also the capability to perform a
full waveform sampling (WF) of the OM signal in addition to the charge measurement
of the PMT pulse and its arrival time. Although this functionality
is mainly used to record double pulses or large
amplitude signals, it is useful to cross-check the computation of the
SPE charge by the integrator circuit of the ARS. In WF mode, 128
digitisations of the OM anode signal are provided, at a sampling rate
of 640 MHz. In order to obtain a precise time stamping of the WF data,
a synchronous sampling of the 50~MHz internal ARS clock is also
performed and read out in addition to the OM data. An example of a WF
record is shown in figure~\ref{AVC} (left). Figure~\ref{AVC} (middle)
displays the charge distribution of the OM signals
obtained by integrating the WF samples after baseline subtraction. The
single photo-electron peak is clearly identified well above the
electronics noise. Figure~\ref{AVC} (right) also shows as a comparison
the charge distribution measured by the Analogue-to-Voltage converter
(AVC) circuit of the ARS. 

\begin{figure}[htb]
\begin{center}
\epsfig{file=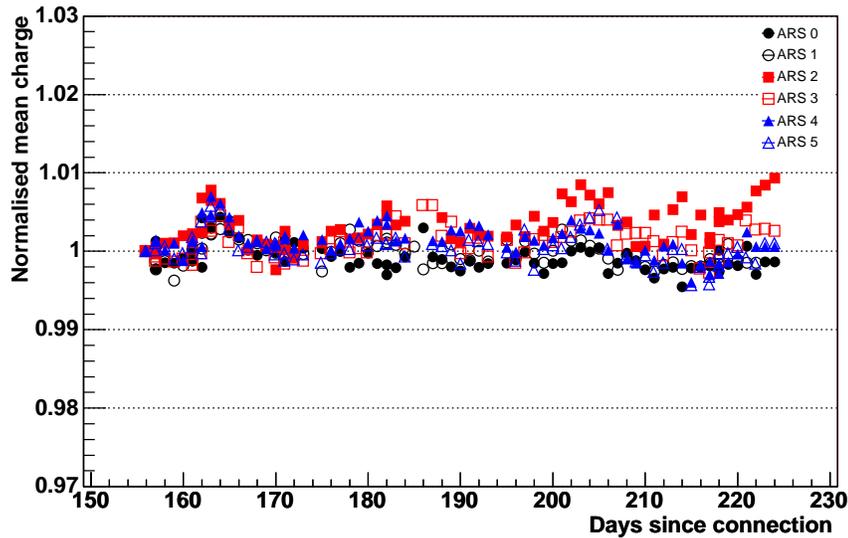,width=0.9\textwidth}
\caption{Variation of the mean charge measured by the Analogue-to-Voltage
converter (AVC) of every ARS as a function of time for the
three OMs. The normalised mean charge is computed as the ratio between
the daily mean value of the charge distribution and the mean value of
their first measurement appearing in the plot.}
\label{Qmean}
\end{center}
\end{figure}

The stability of the measurements of the signal charge obtained with
the integrator circuit of the ARS is shown in figure~\ref{Qmean}. This
figure displays the variation of the mean charge measured by every ARS
AVC circuit as a function of time for the three OMs. The normalised
mean charge is computed as the ratio between the daily mean value of
the charge distribution and the mean value of their first measurement
appearing in the plot. This figure is obtained from the analysis of
minimum bias data largely dominated by SPE events. As can be seen in
figure~\ref{Qmean}, the charge distributions remain stable within
$\pm1$\% for all ARSs during more than two months of monitoring. This
indicates that the PMT gain, the trigger threshold and the charge
integrators did not fluctuate by more than a few percent during that
period.

\section{Optical Module timing precision}

The ANTARES neutrino telescope is designed to have an angular
resolution of less than $0.3^\circ$ for neutrino energies in excess of
10~TeV, which relies on good positioning accuracy (see section~\ref{pos})
and good timing resolution of the signals recorded
by the Optical Modules. The specification for the timing resolution is
such that it should be limited by the transit time spread of the
PMTs which have $\sigma\sim1.3$~ns and by the effect of scattering and
chromatic dispersion of the light during its transmission in water,
which will contribute with a similar amount to the time
uncertainty. To achieve this specification, all electronics and calibration systems
are required to contribute less than 0.5~ns to the overall timing
resolution.

\begin{figure}[bht]
\begin{center}
\epsfig{file=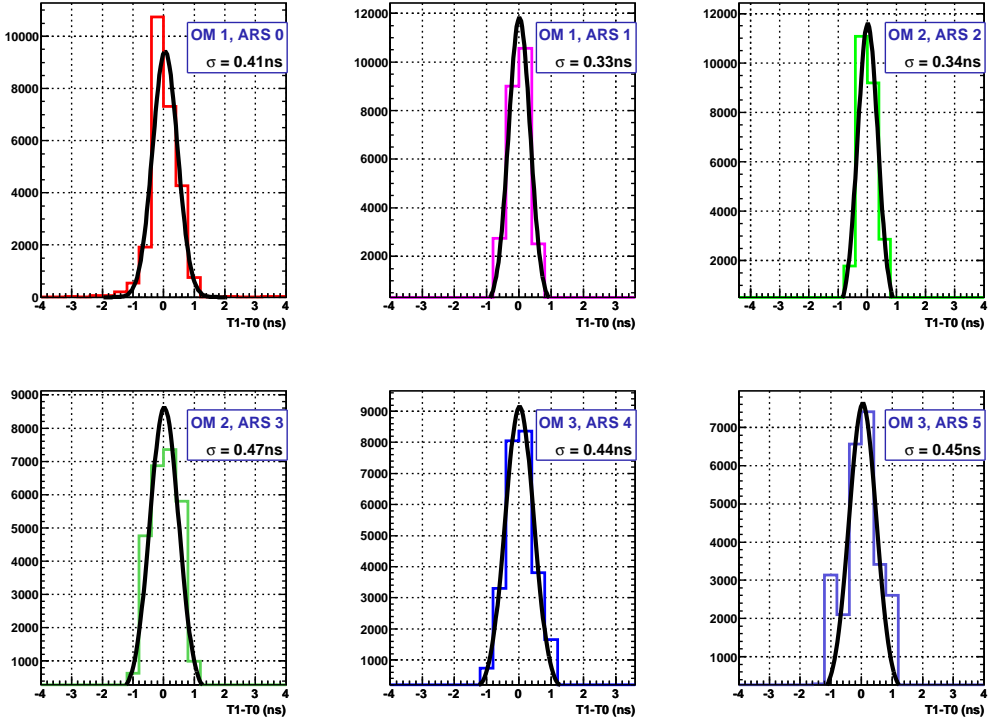,width=1.0\textwidth}
\caption{Distribution of signal arrival times in the Optical Module
(T1) with reference to the time of the LED Optical Beacon flash
(T0). The measurement obtained by every ARS readout channel of the OM
triplet is shown. Each distribution has been centered to 0 by
subtracting its mean value before the fit of a Gaussian function. The
resulting resolution value ($\sigma$) is indicated in each panel.}
\label{LedBeacon}
\end{center}
\end{figure}

The complete timing resolution of the Optical Modules has been
measured with the MILOM using the LED Optical Beacon
system. As indicated in figure~\ref{MILOM
layout}, an LED beacon is located in the bottom storey underneath the
OM triplet of the second storey at a distance of about 15~metres. This
beacon contains 36 individual blue LEDs ($\lambda = 470$~nm,
dominant wavelength) synchronised in time and arranged to give
a quasi-isotropic light emission. A small PMT internal to the LED
beacon monitors the output light pulse timing and amplitude and
provides the time reference of the light flash.

The measurement of the OM timing resolution is performed by pulsing
the LED beacon at a frequency of 30~Hz. Figure~\ref{LedBeacon} shows
the distribution of signal arrival times in the three OMs relative to
the reference PMT in the LED Beacon. For every OM, the measurement
obtained by both ARS readout channels are shown, since every front-end
chip can potentially induce a different intrinsic electronics
contribution to the timing. In the time distribution shown
in figure~\ref{LedBeacon}, the contribution of the small PMT inside
the LED beacon is small since it has a fast rise time of 0.8~ns, and
so the measurement is dominated by the OM contribution. As can be seen
from the distributions shown in figure~\ref{LedBeacon}, the timing
resolution of all Optical Module readout channels is measured to be
$\sigma\sim0.4$~ns. This resolution is however obtained for large intensity
light pulses and so is not dominated by the PMT transit time spread
but by the intrinsic electronics resolution. A detailed analysis of the
dependence of timing resolution on light intensity is required to
separate the various contributions, but this result already shows that
the complete electronics contribution is smaller than 0.5~ns as
required.

\begin{figure}[htb]
\begin{center}
\epsfig{file=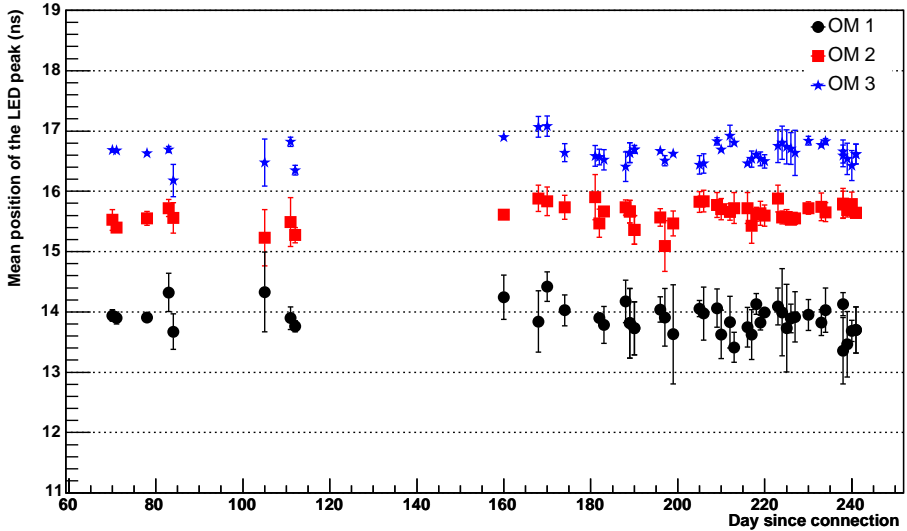,width=1.0\textwidth}
\caption{Mean values of the arrival times of the OM signals induced by
an internal LED flash with respect to the time of the flash. The mean
time is displayed with reference to the start of the Time-to-Voltage
converter (TVC) of the ARS. The plot displays the measurements
obtained for the three OMs over a six month period. For every OM,
the data points
display the average value measured by its two ARS channels.}
\label{TTled}
\end{center}
\end{figure}

Every Optical Module also contains an internal blue LED in order to
monitor the stability of the photomultiplier tube. This LED is mounted
on the back of the PMT, as shown in figure~\ref{figOM}, and
illuminates a large fraction of the photocathode through the
phototube. The transit time of the PMT is monitored by flashing the
internal LED at a rate of about 100~Hz and by looking at the OM signal
arrival time with respect to the time of the LED
flash. Figure~\ref{TTled} shows the average values of these delays for
each of the three OMs as a function of time. The results confirm that
the transit time of each OM has remained stable to within 0.5~ns,
throughout the six month period considered.

\section{Acoustic positioning system resolution}
\label{pos}

The second essential element to achieve the necessary angular
resolution of the neutrino telescope is a real time measurement of the
position in space of the Optical Modules with a precision of
$\sim10$~cm. These positions are obtained by triangulation using
distance measurements provided by the acoustic positioning system. The
full acoustic positioning system will consist of a three dimensional
array of emitting transducers (RxTx modules) fixed in known positions on the sea bed,
together with receiving hydrophones (Rx modules) attached on several
storeys along every detector line. These devices exchange precisely
timed acoustic signals in the 40-60~kHz frequency range. In addition,
the acoustic system includes four autonomous transponders which emit
an acoustic ``ping'' at a precisely given frequency in response to
a special interrogation by one RxTx module. These transponders will be
located around the ANTARES detector in order to enlarge and make
more uniform the geometry of the triangulation basis. At the
present time, only a limited number of acoustic devices are installed
at the site: one RxTx module on the MILOM anchor and one Rx module on
its first storey. A first autonomous transponder has also been
installed at the ANTARES site, its transducer being fixed at about 4 m
above the sea bed on top of a pole supported by a pyramidal structure
at a horizontal distance of $\sim175$~m from the MILOM anchor.

\begin{figure}[htb]
\begin{center}
\epsfig{file=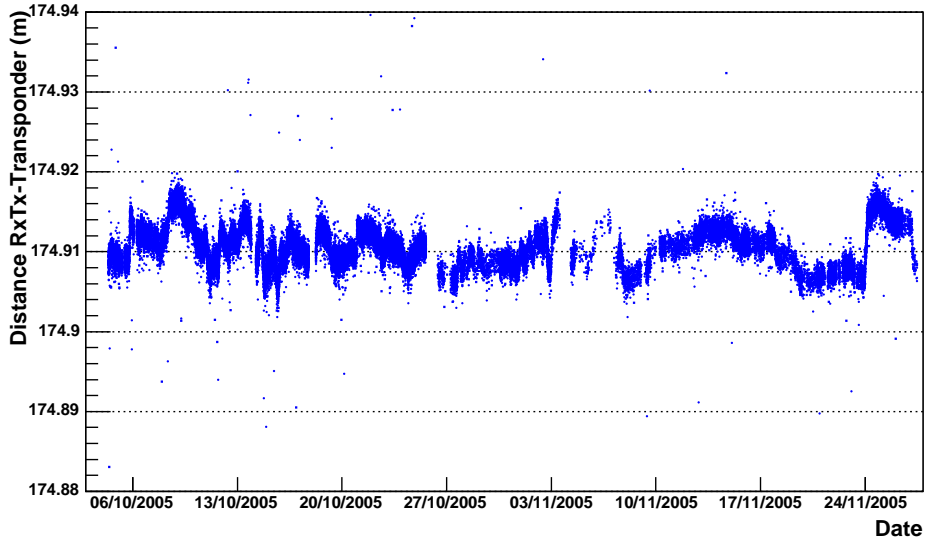,width=1.0\textwidth}
\caption{Distance measured by the acoustic positioning system between
the RxTx module and the autonomous transponder, both being fixed a
few metres above the sea bed. The measurement is displayed as a
function of time, during two months of operation.}
\label{Acoustic}
\end{center}
\end{figure}

Although the limited number of installed acoustic emitters does not
allow the triangulation reconstruction, the MILOM operation
demonstrates the resolution and the stability of the acoustic system
by monitoring the distance measured between two fixed points. This is
illustrated in figure~\ref{Acoustic} which displays the acoustic
distance measured between the RxTx transducer attached on the MILOM
anchor and the autonomous transponder. The acoustic system
measurements show a resolution of a few mm and a stability of
$\sim1$~cm over a distance of 174.91~m, during two months of
operation. The absolute distance obtained by this system is well in
agreement with the value of 175.6~m obtained with the long base line
acoustic navigation system used onboard the surface boat during the
marine operation to monitor the deployment and installation of the
lines. This navigation system, based on 10~kHz range acoustic signal
exchanges, allows range measurements of several thousands of metres
with an accuracy limited to $\sim1$~m. With the MILOM operation, the
performance of the acoustic positioning system is confirmed to be
well within the specification required to obtain a precision of
$\sim10$~cm on the spatial reconstruction.

\section{Examples of instrumentation measurements}

\begin{figure}[htb]
\begin{center}
\epsfig{file=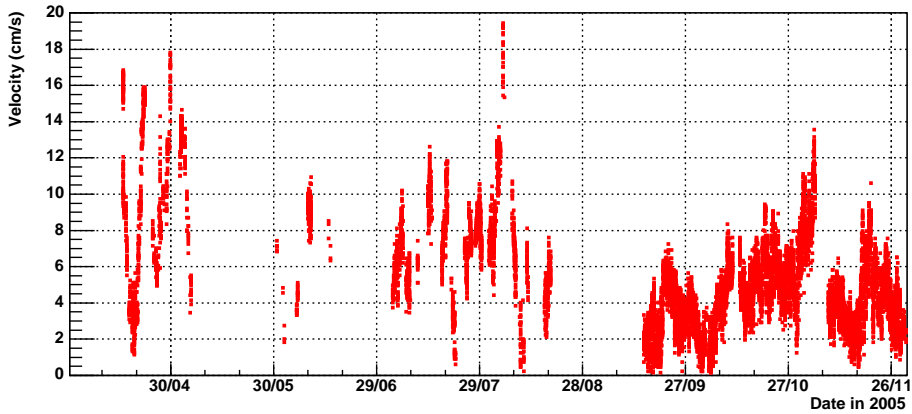,width=1.0\textwidth}
\caption{Velocity of the water current flow measured by the Doppler
current profiler, as a function of time over a period of eight months.}
\label{Current}
\end{center}
\end{figure}

In parallel to the Optical Module data taking and to the operation of
the calibration devices, the various instruments of the MILOM
dedicated to environmental measurements have been regularly read
out. This has allowed a continuous monitoring of some physical quantities
of the sea water at the ANTARES site which might have an influence on
the calibration of the detector or on the bioluminescence background:
the water current flow; the water temperature; the sound velocity and
the water transparency. As an example, the measured water current
velocity recorded by the Doppler current profiler located on the MILOM
top storey is displayed in figure~\ref{Current}, as a function of time
over a period of eight months. These measurements confirm that the
water current usually remains small at the ANTARES site, with a
velocity not exceeding 20~cm/s and with an average speed of
$\sim5$~cm/s.

\begin{figure}[htb]
\begin{center}
\epsfig{file=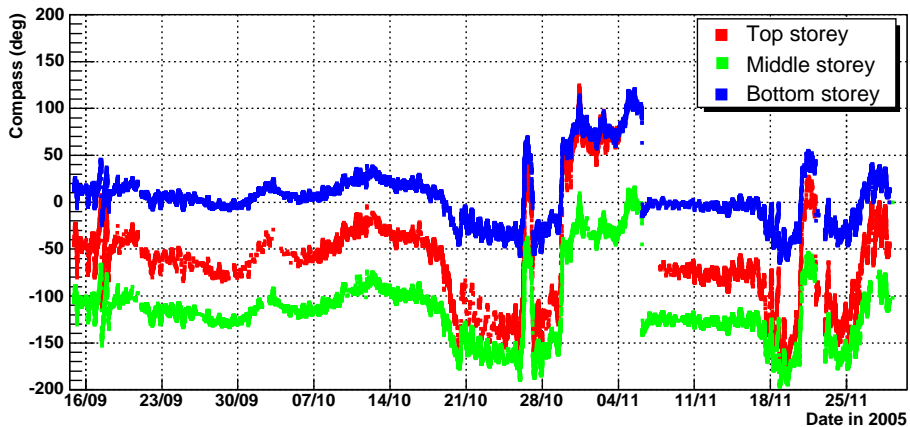,width=1.0\textwidth}
\caption{Individual headings of the three MILOM storeys, measured by
the compass included in every electronics container, as a function of
time during two months of monitoring.}
\label{Compass}
\end{center}
\end{figure}

Figure~\ref{Compass} shows the individual headings of the three
storeys of the line measured by the compass in each electronics
container, as a function of time for the same period. Although each
storey has its own relative orientation due to the uncontrolled
alignment in construction, the rotation changes affect, in general,
the whole line. As expected, the top storey
shows a larger rotation amplitude than the others. The
measurements also indicate that the two lower storeys, separated by a
shorter cable of 12.5~m, tend to remain at a constant orientation
with respect to each other. Some correlation between the storey
rotations and the water direction changes can also be noticed,
especially when the current velocity is large as during the first
days of November 2005.

The seismometer read out by the MILOM has been delivering continuous
data to the ANTARES shore station. It provided clear signals from
several earthquakes around the world, such as the one in Japan on
August $30^{th}$ 2005 and in Peru on September $24^{th}$ 2005. Further
analyses of the seismometer data are in progress.

\section{Conclusions}
Since April 2005, the ANTARES Collaboration has been operating a Mini
Instrumentation Line equipped with Optical Modules immersed at
a depth of 2500~m on the ANTARES site. The main purpose of this line,
built with pre-production elements of the detector, was to
allow an in-situ check of all the equipment and in particular to
validate the performance of the time calibration and acoustic
positioning devices.

The various data regularly acquired during the MILOM operation confirm
the capability of the Optical Modules and of its front-end electronics
to trigger and read out single photo-electron signals. The continuous
data collection from the OMs and the MILOM instruments during several
months validate the whole electronics and the DAQ readout system
designed for the detector. The operation of the light source
calibration devices, such as the LED Optical Beacon or the OM internal
LED, shows that a time calibration of all Optical Modules can be
achieved in-situ with an accuracy better than 0.5~ns. The results
yield an electronics contribution to the OM timing resolution smaller
than 0.5~ns, as required. Finally, the operation of the acoustic
positioning system shows that its performance is well within the
specification required to achieve a space resolution of
$\sim10$~cm. All these results confirm the good performance of
the electronics and calibration devices as required
to obtain the desired angular resolution of the detector.
The experience with the MILOM gives confidence for the
operation of the first complete lines of the ANTARES neutrino telescope,
the deployment of which has started in early 2006.

\section*{Acknowledgements}
The authors acknowledge the financial support of the funding agencies. In
particular: Centre National de la Recherche Scientifique (CNRS), Commissariat
\`a l'\'Energie Atomique (CEA), Commission Europ\'eenne (FEDER fund), R\'egion
Alsace (contrat CPER), R\'egion Provence-Alpes-C\^ote d'Azur,
D\'epartement du Var and Ville de La Seyne-sur-Mer, in France;
Bundesministerium f\"ur Bildung und Forschung (BBF), in Germany; Istituto
Nazionale di Fisica Nucleare (INFN), in Italy; de stichting voor Fundamenteel
Onderzoek der Materie (FOM) and the Nederlandse organisatie voor
Wetenschappelijk Onderzoek (NWO), in the Netherlands; Russian
Foundation for Basic Research (RFBR), in Russia; Ministerio de Educaci\'on y
Ciencia (MEC), in Spain.

\end{document}